# Derived Parameters for NGC 6791 from High-Metallicity Isochrones

Michael J. Tripicco and R. A. Bell

Astronomy Department, University of Maryland, College Park, MD 20742-2421

electronic mail: miket@astro.umd.edu, rabell@astro.umd.edu

Ben Dorman

Department of Astronomy, University of Virginia, Charlottesville, VA 22903-0818[1]

electronic mail: dorman@shemesh.gsfc.nasa.gov

and

Beth Hufnagel

Board of Studies in Astronomy & Astrophysics, UC Observatories/Lick Observatory,
University of California, Santa Cruz, CA 95064

and

National Radio Astronomy Observatory, PO Box 0, Socorro, NM 87801[2]

electronic mail: bhufnage@aoc.nrao.edu

## ABSTRACT

We have computed 8, 10, and 12 Gyr isochrones and physically consistent models of zero-age red horizontal branch stars for stellar masses between 0.55 and 1.3 $M_\odot$, all at [Fe/H] = +0.15. Comparison to the NGC 6791 BVI photometry of Kałużny & Udalski (1992, Acta Astron., 42, 29) and Montgomery, Janes & Phelps (1994, AJ, 108, 585) yields an age of $10.0 \pm 0.5$ Gyr at an apparent distance modulus $13.49 < (m - M)_V < 13.70$. The color offsets required to fit the isochrones, combined with the spectroscopic results of Friel & Janes (1993, A&A, 267,75), imply that the foreground reddening to NGC 6791 lies in the range $0.24 > \mathrm{E}(B - V) > 0.19$ with $+0.27 < [\mathrm{Fe/H}] < +0.44$. These results are derived using a technique by which we predict color shifts and apply these to the isochrones to simulate progressively higher metallicities.

The zero-age horizontal branch models suggest that the red horizontal branch stars of NGC 6791 have masses $\lesssim 0.7 M_\odot$. The masses are similar to those found for M67 red horizontal branch stars by Tripicco, Dorman & Bell (1993, AJ, 106, 618) and for globular cluster red horizontal branch stars, even though the M67 progenitors are $\sim 0.2 M_\odot$ more massive, while the progenitors of globular cluster horizontal branch stars are similarly less massive. This suggests the presence of a mechanism, not strongly dependent on metallicity, which reduces stellar envelopes on the zero-age horizontal branch *to* a given mass rather than *by* a given amount.

*Subject headings:* clusters: open – clusters: individual (NGC 6791) – stars: evolution

## 1. Introduction

The rich open cluster NGC 6791 is one of the oldest and most metal-rich objects of its class, and its study can therefore provide important constraints on models of galactic chemical and dynamical evolution. In recent years a number of groups have obtained extensive high-quality photometry for this cluster, most notably Kałużny & Udalski (1992; hereafter KU92) and Montgomery, Janes & Phelps (1994; hereafter MJP94) The cluster metallicity is clearly at least solar and probably higher, with the best recent determination being [Fe/H] = $+0.19 \pm 0.19$ (Friel & Janes, 1993) based on spectroscopy of nine giants. Yet the absence of greater than solar metallicity isochrones in the literature has forced most groups to fit their data with isochrones computed for [Fe/H] = 0 and then extrapolate to higher metallicity to determine the cluster parameters. The result has been that large uncertainties remain in the age, reddening and distance to NGC 6791.

As part of an ongoing project to generate synthetic integrated spectra for globular clusters and galaxy nuclei, we have calculated new isochrones and horizontal branch evolutionary tracks over a wide range of metallicities and ages. Our most metal-rich isochrones have a metal mass fraction (Z) of 0.024 (compared to 0.0169 for the Sun; see Sec. 2.1 below) or [Fe/H] $\approx$ +0.15. Given the closeness of the metallicity of these isochrones to the value cited above for NGC 6791, one can expect that they should yield the most reliable estimates of the cluster parameters to date. If, as we shall argue, the metallicity of NGC 6791 is higher than this, we can estimate the effect of a small metallicity increase on the colors of the isochrones by examining a few test models run at higher values of Z.

This paper presents these calculations and demonstrates their application to the color-magnitude diagram (CMD) of NGC 6791. We show that the offset between the theoretical isochrones and the data cannot be entirely due to the reddening, but must also contain a component due to metallicity. We find that in order to satisfy both the photometric and spectroscopic observational constraints, the metallicity must lie in the range $+0.27 <$ [Fe/H] $< +0.44$, and the reddening $0.24 >$ E$(B - V) > 0.19$.

The format of the paper is as follows: Sec. 2 describes the theoretical calculations. The isochrones are compared with published cluster BVI photometry in Sec. 3 and the cluster parameters are determined. In Sec. 4 we compare our results with those of other studies and Sec. 5 summarizes the conclusions.

## 2. Computational Aspects

All of the calculations which lead to the present isochrones have been performed at the University of Maryland using a variety of computer programs which follow stellar evolution, compute model atmospheres, generate synthetic stellar spectra and finally convolve these with filter transmission functions to yield calibrated magnitudes and colors. We note that the process used here is virtually identical to that described for the solar abundance case in Tripicco, Dorman & Bell (1993; hereafter TDB93).

For consistency with previous calculations we continue to use the Los Alamos (Huebner et al. 1977) and Alexander (1975; 1981) opacities in the present stellar evolution calculations. Note that low temperature Alexander tables are known to overestimate the H$^-$ opacity (Alexander, Augason, & Johnson 1989) and thus underestimate the surface temperatures of the models. The effect of this is to change the value of the convective mixing length parameter, $\alpha$, that reproduces the solar properties. A metal mass fraction $Z = 0.024$ ([Fe/H] $\approx +0.15$), helium mass fraction $Y = 0.268$ and $\alpha = 1.6$ were used throughout. Our choices for Y and $\alpha$ were guided by running evolutionary tracks for the Sun, with $Z_\odot = 0.0169$. This solar Z is somewhat lower than the commonly adopted value of 0.02 (see Sackmann, Boothroyd & Fowler (1990) for a listing of values used by different authors) and so it we briefly digress to consider this point in more detail.

## 2.1. Standard Solar Model

The solar abundances which we used for H, He, C, N, O, Ne, Na, Mg, Al, Si, S, K, Ca, Cr, Fe and Ni are given in the column labeled "SSG/MARCS" in Table 1. These values are the ones used by the MARCS model atmosphere program (Gustafsson et al., 1975) and the SSG synthetic spectrum program (Bell & Gustafsson, 1978, 1989; Gustafsson & Bell, 1979). In addition to those listed above, Table 1 contains the abundances of some additional elements, e.g. Sc, Ti, V, Cr, Mn and Co, which are important contributors to the line absorption in cool stars. SSG does, however, use the abundances of nearly all elements in the periodic table. In general, the abundances shown in Table 1 are ones which we have used for some time; they correspond to mass fractions X= 0.7038, Y= 0.2795, Z= 0.0168. This Z value is consistent with the opacity table we have used for all solar evolutionary track calculations to date (Z= 0.0169). The last column in Table 1 indicates for comparison the abundances from Anders & Grevesse (1989, hereafter AG89). The full slate of abundances from AG89 yields X= 0.7065, Y= 0.2742, Z= 0.0194. Much of the difference between the SSG and AG89 Z values (i.e., 0.0014 out of 0.0026) comes from the more recent value for the oxygen abundance which we use. The value of $8.86 \pm 0.04$, found by Biémont et al. 1991, is identical to what we have used for some years while the value quoted by AG89 is the older value of 8.93 derived by Lambert (1978). We note that the C and N abundances of $8.57 \pm 0.04$ and $7.99 \pm 0.04$ quoted by Biémont et al. (1993) are very similar to the ones we use, as are recent values of the Fe abundance, e.g. $7.48 \pm 0.09$ (Holweger, Heise & Koch, 1990) and $7.50 \pm 0.07$ (Holweger et al., 1991).

Given Z, we then determined the values for the helium abundance and mixing length for which a 1.0 $M_\odot$ evolutionary track reaches the solar luminosity and effective temperature at the solar age. We found that $Y = 0.2693$ and $\alpha = 1.424$ produced the single best model, passing through $L/L_\odot = 1.0$ at an age of 4.54 Gyr with $T_{\text{eff}} = 5778$ K and log g= 4.438. However, in TDB93 we showed that our tracks best fit the RGB of the solar abundance open cluster M67 when $\alpha = 1.6$. For maximum consistency with that case, and because the fit to the RGB is incidental to the determination of the cluster age and reddening we chose to stay with $\alpha = 1.6$. Because the age at which the evolutionary track passes through $L/L_\odot = 1.0$ is much more sensitive to Y than to $\alpha$, we were able to obtain an acceptable solar model using $\alpha = 1.6$ if the helium fraction was slightly lowered to Y= 0.2680. These parameters were adopted for all of the evolutionary tracks discussed in the present paper. The use of the new opacities in the future will likely lead to different values for the solar Y and $\alpha$ but we consider it unlikely that the overall results for NGC 6791 determined via this analysis will be significantly affected.

Table 1. Solar Abundances

| Element | SSG/MARCS | AG89[a] |
|---|---|---|
| H  | 12.00 | 12.00 |
| He | 11.00 | 10.99 |
| C  | 8.62  | 8.56  |
| N  | 8.00  | 8.05  |
| O  | 8.86  | 8.93  |
| Ne | 7.70  | 8.09  |
| Na | 6.46  | 6.33  |
| Mg | 7.44  | 7.58  |
| Al | 6.51  | 6.47  |
| Si | 7.68  | 7.55  |
| S  | 7.39  | 7.21  |
| K  | 5.06  | 5.12  |
| Ca | 6.54  | 6.36  |
| Sc | 3.20  | 3.10  |
| Ti | 4.78  | 4.99  |
| V  | 4.10  | 4.00  |
| Cr | 5.70  | 5.67  |
| Mn | 5.40  | 5.39  |
| Fe | 7.48  | 7.67  |
| Co | 4.70  | 4.92  |
| Ni | 6.30  | 6.25  |

[a]Anders & Grevesse 1989

### 2.2. Calculation of Isochrones

First, evolutionary tracks for stars having masses between 0.4 to 1.5 $M_\odot$ were computed using the stellar evolution code STEV (Dorman 1992). This program is a rewritten version of the University of Victoria code described by VandenBerg (1983; 1992) which has been augmented by a detailed treatment of the helium-burning evolutionary phases. For each mass we follow the evolution from the zero-age main sequence (ZAMS) through central hydrogen exhaustion to the red giant branch (RGB) which terminates at the He-flash. In addition, we have generated zero-age horizontal branch (ZAHB) models for masses between 0.55 and 1.3 $M_\odot$ (using a core mass of 0.467 $M_\odot$ determined from terminal RGB models).

As discussed in TDB93 and by VandenBerg (1992), the position and slope of the RGB at a given mass and metallicity depends rather sensitively on a number of poorly constrained parameters, notably $\alpha$ and the surface pressure boundary condition. Errors in the Alexander (1981) low-temperature opacities led VandenBerg to apply empirically determined corrections to the model-atmosphere-based pressure tables (which are used to set the surface pressure boundary condition) in order to produce RGBs whose slopes agree with IR color-temperature relations for globular clusters. These adjustments to the tables were used in the present calculations (as in TDB93) essentially for cosmetic purposes; the fit to the main-sequence turnoff, where the model calculations are based strictly on MARCS surface pressures, and the brightness of He-burning clump stars are primarily what are used to determine the cluster age, reddening and distance.

The evolutionary tracks (ZAMS to He-flash) have been transformed into isochrones for ages of 8, 10 and 12 Gyr using the usual technique of equivalent-evolutionary points (Prather 1976; Bergbusch & VandenBerg 1992; Dorman, O'Connell, & Rood 1995).

In the next step we generate synthetic spectra for stellar models at intervals along the isochrones, as well as on the ZAHB. At each selected ($T_{\text{eff}}$ – log g) point a plane-parallel, flux-constant model atmosphere was calculated using MARCS, which was then used as input to compute a synthetic spectrum with the SSG code. The spacing used between flux points was 0.1 Å, and all of the spectra were computed between 3000 and 12000 Å. We point out that for a variety of reasons, models for stars cooler than about 4000 K are still quite uncertain (see Paltoglou & Bell 1994). One of the main difficulties arises from the dominance and strong temperature sensitivity of the TiO absorption bands. Work is in progress to update the handling of molecular species in the SSG program; for the present paper we have synthesized all of the spectra without using TiO.

For the final step in the process we convolve each synthetic spectrum along the isochrones with filter transmission profiles to determine the colors. Though we concentrate here on the broadband BVI colors (using profiles from Bessell 1990), simulated photometry for other color systems (e.g., DDO, Washington) has also been carried out. The colors have been calibrated in a model-independent manner using the Gunn & Stryker (1983) spectrophotometric scans as described in Tripicco & Bell (1991). Absolute magnitudes and bolometric corrections were determined relative to our standard SSG solar model (with $T_{\mathrm{eff}} = 5770$ K and $\log g = 4.44$) for which we adopt the values $M_{V_\odot} = +4.83$ and $\mathrm{B.C.}_\odot = -0.07$.

Because of the volume of data, we have chosen not to include a table of the isochrones. They are, however, available upon request from the first author.

## 3. Analysis

### 3.1. Fitting Isochrones to NGC 6791

We now compare our theoretical isochrones and ZAHB models with two independent sets of photometric data for NGC 6791: from KU92 and MJP94. Both groups obtained their data using the same telescope and instrumental setup. The full set of data from KU92 contains BVI photometry for 7115 stars near the cluster center but we have chosen to focus on a small subset of their data in this paper: 445 stars whose membership probability have been determined via proper-motion study to be 50% or higher (Cudworth 1994). The same proper-motion results were used by MJP94 to select 495 high-probability cluster members from their data (see their Fig. 11). In both cases we felt that using these subsets of the full photometric samples would result in the most accurate determination of the cluster parameters. We have, however, confirmed in each case that they would also apply to the full sets of data for the central region of the cluster.

In Fig. 1 we compare the KU92 $M_V$ vs. $(B-V)$ CMD with our isochrones. If we adopt an apparent distance modulus $(m-M)_V = 13.52$ and shift the isochrones in color by $\Delta(B-V) = 0.25$, the data are well-matched by our 10 Gyr isochrone, particularly in the critical regions around the main-sequence turnoff (MSTO), the subgiant branch (SGB) and the unevolved main-sequence. Also, with this choice of parameters, the ZAHB model sequence forms a natural lower boundary to the clump of He-burning stars as appropriate for post-ZAHB evolution (see Fig. 3 of TDB93). As is well-known, the magnitude difference between the MSTO and the ZAHB is extremely sensitive to age. The excellent fit to both the clump and the turnoff stars thus strongly constrain the age and apparent distance modulus of the cluster. The color offset $\Delta(B-V)$ represents an upper limit to the cluster reddening: we will return to this important point below.

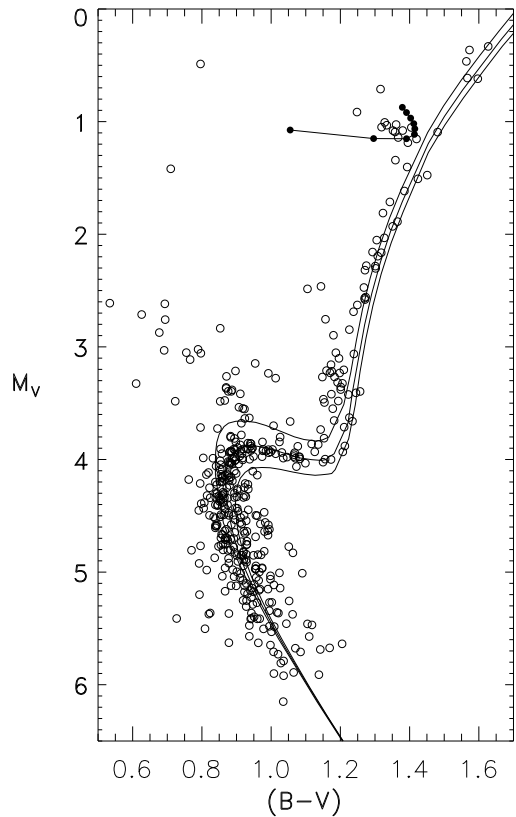

Fig. 1.— $M_V$ vs. $(B-V)$ color-magnitude diagram for 445 stars whose probability of membership in NGC 6791 have been determined to be greater than 50% via proper motion studies. The photometry was drawn from the sample of KU92. The lines represent our theoretical 8, 10, and 12 Gyr isochrones and associated zero-age horizontal branch for [Fe/H] = +0.15. Filled circles on the latter denote models for masses between 1.30 and 0.55 $M_\odot$. The model sequences have been shifted by $\Delta(B-V) = 0.25$ mag. and the data by $(m-M)_V = 13.52$ mag.

Fig. 2 presents a similar comparison between the isochrones and the MJP94 $M_V$ vs. $(B-V)$ CMD. The data seem considerably tighter in this case, and once again the 10 Gyr isochrone fits quite well, as does the ZAHB sequence. The offsets are slightly different, with $(m-M)_V = 13.49$ and $\Delta(B-V) = 0.275$ providing the best fit. These discrepancies are consistent with the small systematic differences between the two datasets as computed by MJP94 and illustrated in their Fig. 4.

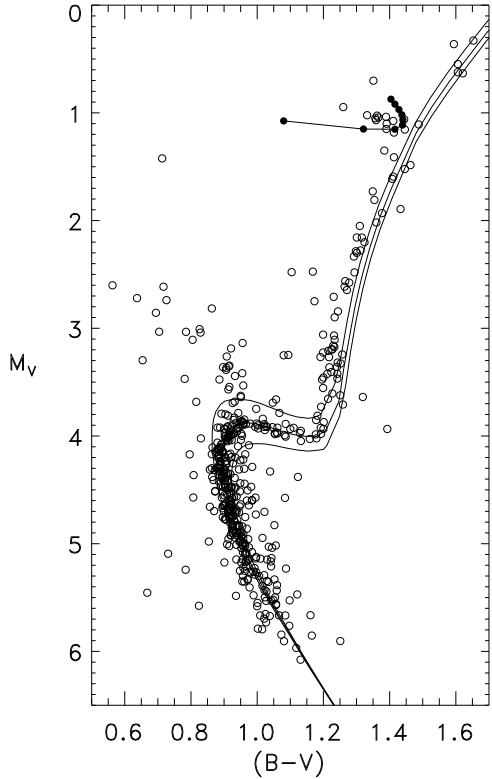

Fig. 2.— Same as Fig. 1 except that the photometry (for 495 high-probability cluster members) comes from MJP94. In this case $\Delta(B-V) = 0.275$ mag. and $(m-M)_V = 13.49$ mag.

We now turn to the $M_V$ vs. $(V-I)$ plane. Fig. 3 shows the photometry from KU92 and Fig. 4, from MJP94. There are several features to note about these comparisons. One is that in these diagrams it is the KU92 data which form a tighter sequence, whereas for $(B-V)$ it was just the opposite. In both cases the 10 Gyr isochrone fits the MSTO, SGB and lower RGB quite well. And in both the fit deteriorates towards redder $(V-I)$ colors. For the brightest giants some of this is undoubtedly due to the effects of TiO absorption which has been neglected in the models. Uncertainties in the calibration of the model colors may also play a role, owing to problems in correcting for telluric absorption in the region of the I passband in the Gunn & Stryker (1983) scans. In any event, the values of the color offset, $\Delta(V-I)$, are 0.315 and 0.370 for KU92 and MJP94, respectively. The difference between the two datasets is considerably larger than was the case for $(B-V)$. In their Fig. 4, MJP94 show that their $(V-I)$ colors average 0.03 magnitudes redder than KU92. However, that refers only to stars brighter than V magnitude 17, which is somewhat brighter than the turnoff. As one continues toward fainter magnitudes (where the isochrone fit is primarily determined) the average residuals between the two samples as illustrated by their diagram increases on average, making it plausible that the 0.055 magnitude discrepancy between the $\Delta(V-I)$ offset values arises from systematic differences in the I-band observations.

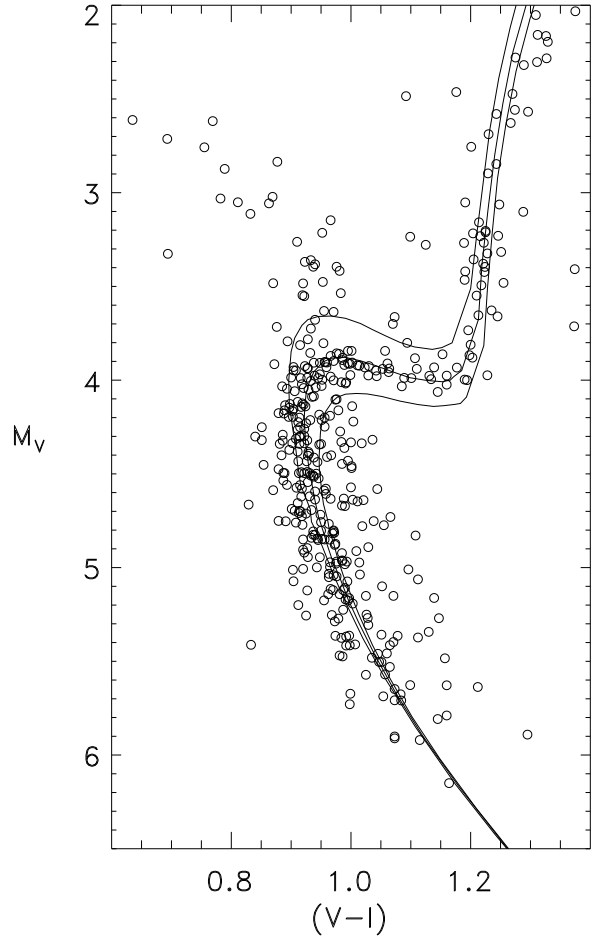

Fig. 3.— $M_V$ vs. $(V-I)$ color-magnitude diagram for the same 445 stars from KU92 as in Fig. 1. The model color offset is $\Delta(V-I) = 0.315$.

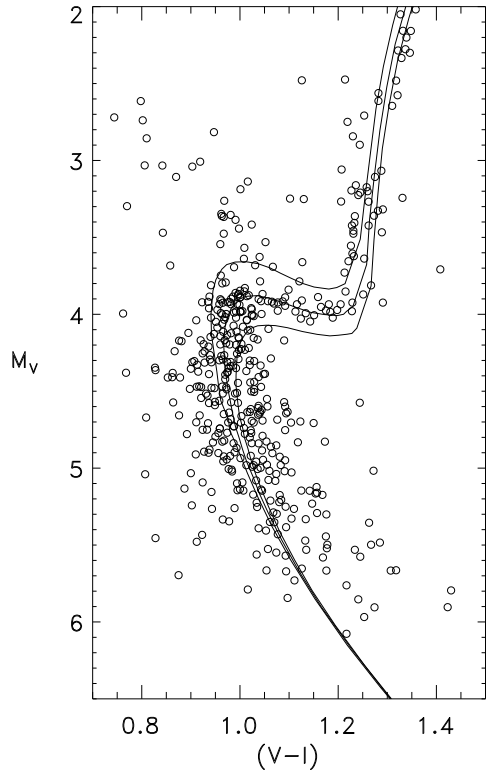

Fig. 4.— Same as Fig. 3, but for the MJP94 photometry. The isochrones have been offset by $\Delta(V-I) = 0.37$.

### 3.2. Decoupling the Effects of Reddening and Metallicity

There are several pieces of evidence which suggest that the substantial $\Delta(B-V)$ offset needed to match the present isochrones to the CMD of NGC 6791 has two components, i.e.,

$$\Delta(B-V) = \mathrm{E}(B-V) + \delta(B-V)$$

Here, $\mathrm{E}(B-V)$ is the actual foreground reddening, while $\delta(B-V) > 0$ is an additional shift that arises because the "true" cluster metallicity is greater than that of the isochrones ([Fe/H] $\approx +0.15$). Any systematic error present in our theoretical colors would, of course, add a third component to $\Delta(B-V)$; neglecting this would cause the derived reddening to be correspondingly overestimated. However, we believe that any errors in our theoretical $(B-V)$ colors are quite small (at least for metallicities as high as solar and for stars hotter than about 4000 K) based on comparisons between observed and calculated colors for particular stars (Bell, Paltoglou & Tripicco, 1994).

The first argument which suggests that $\Delta(B-V) > \mathrm{E}(B-V)$ is that despite significant variations in the derived cluster reddening, no value for $\mathrm{E}(B-V)$ as high as 0.25 or 0.275 has ever been reported. Most conventional techniques have generally centered around numbers between 0.1 and 0.2. In addition, KU92 derived an upper limit of $\mathrm{E}(B-V) \leq 0.195$ based on the colors of several sdB/sdO candidates in NGC 6791. (A more detailed reanalysis of these stars by Liebert, Saffer & Green 1994 leads to $\mathrm{E}(B-V) \approx 0.14$.)

The second indication is based on the results of the spectroscopic study of NGC 6791 giants by Friel & Janes (1993). Because they used stellar temperatures derived from a $(B-V) \to T_{\mathrm{eff}}$ relation, their results can only be expressed in terms of a coupled set of values. If the foreground reddening were to be higher than the value they adopted, the temperature derived for a given star would be greater and the observed line strengths would imply a higher metallicity. Their original analysis found the mean [Fe/H] for nine stars to be +0.19 for $\mathrm{E}(B-V) = 0.12$. Had they used $\mathrm{E}(B-V) = 0.19$, then the result would have instead been [Fe/H] = +0.28 (see Garnavich et al. (1994)).

Thus, assuming that $\Delta(B-V)$ is entirely due to reddening leads to a contradiction: the spectroscopic evidence would then imply a still higher metallicity value, much greater than that used to calculate the isochrones. If, however, the metallicity of the cluster is greater than that of the isochrones, then the color offset must have the two components described above. For any value of the reddening, there is a value of $\delta(B-V)$ – corresponding to a larger estimate of the cluster metallicity — that implies a corresponding adjustment which should be made to the composition of the isochrones. Further, there is a unique solution which satisfies both the spectroscopic results and the isochrone fitting. That is, suppose we know (1) how the isochrones shift in color with metallicity — the $\delta(B-V)$ vs. [Fe/H] relationship — and (2) how the spectroscopic determination of [Fe/H] varies with the assumed reddening. Then, since $\delta(B-V) = \mathrm{Constant} - \mathrm{E}(B-V)$, we have two curves on the $\mathrm{E}(B-V)$–[Fe/H] plane which have a unique intersection.

First, we need to determine how much the isochrones will shift as the adopted metallicity is increased. We employ a technique similar to that used by MJP94. We have computed ZAMS models for 0.8 $M_\odot$ at various values of $Z$ and with all other parameters fixed. The choice of 0.8 $M_\odot$ is somewhat arbitrary and is designed to be low enough that it can be regarded as unevolved over the age range considered here, yet high enough that it lies on the observable portion of the main sequence. We then determine their colors as described in the previous section. The results are provided in Table 2. Figs. 5 and 6 show the variation of $(B-V)_{(0.8)}$ and $\mathrm{M}_{\mathrm{V}(0.8)}$, respectively, as a function of $Z$ (filled triangles).

Table 2. 0.8 M$_\odot$ ZAMS Models

| Z | M$_V$ | (B−V) | [Fe/H] |
|---|---|---|---|
| 0.0001 | 5.430 | 0.422 | −2.24 |
| 0.0006 | 5.472 | 0.454 | −1.46 |
| 0.0040 | 5.773 | 0.602 | −0.64 |
| 0.0169 | 6.655 | 0.935 | 0.00 |
| 0.0240 | 7.018 | 1.058 | +0.15 |
| 0.0400 | 7.556 | 1.228 | +0.38 |
| 0.0600 | 7.927: | 1.329: | +0.57 |

Note. — Y= 0.268, $\alpha$ = 1.6 throughout. Values for the highest metallicity are slightly less reliable than the rest.

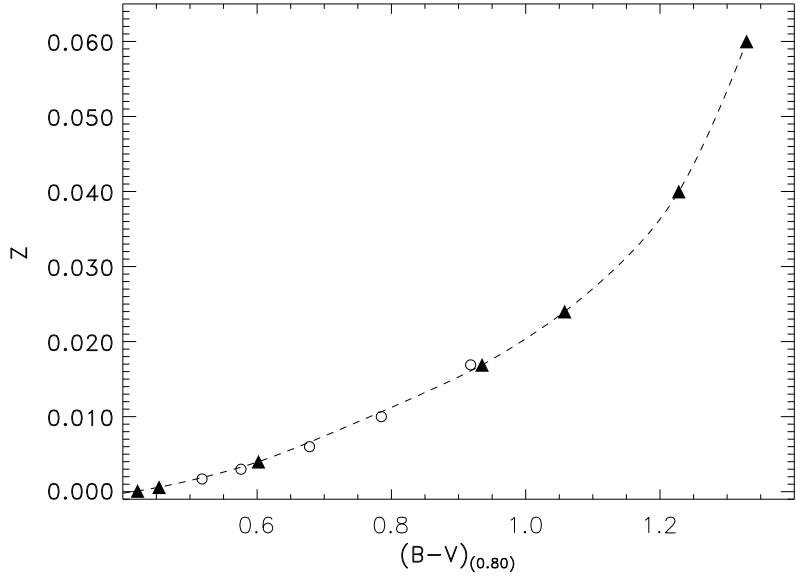

Fig. 5.— $(B-V)$ color as a function of metallicity, Z, for zero-age main-sequence 0.8 $M_\odot$ models. The filled triangles represent our calculations and the open circles are from VandenBerg (1985). Y = 0.268 and $\alpha$ = 1.6 for all of our models. The dashed line is a spline fit used for interpolation between our model points.

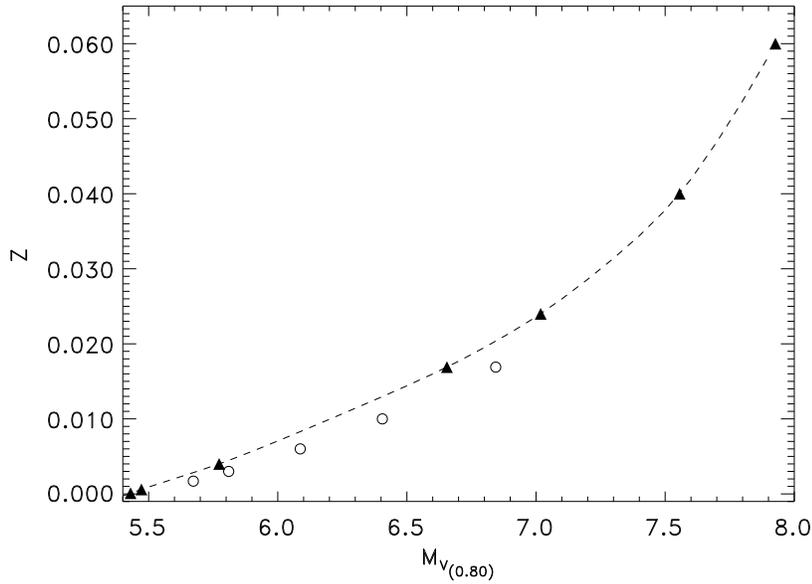

Fig. 6.— Same as Fig. 5, but showing the variation of $M_V$ with Z. The small systematic difference between our calculations and those of VandenBerg (1985) presumably reflect the difference in the adopted helium mass fraction, Y.

For purposes of comparison we also plot the same quantities from VandenBerg (1985, open circles). Our models were calculated with Y= 0.268, while VandenBerg's models have Y= 0.25. This accounts for the fact that our models are slightly brighter at a given Z, as seen in Fig. 6. The smooth $M_{V(0.8)}$ vs. Z and $(B-V)_{(0.8)}$ vs. Z relations for our models have been fit using cubic splines (for later interpolation) as indicated by the dashed lines.

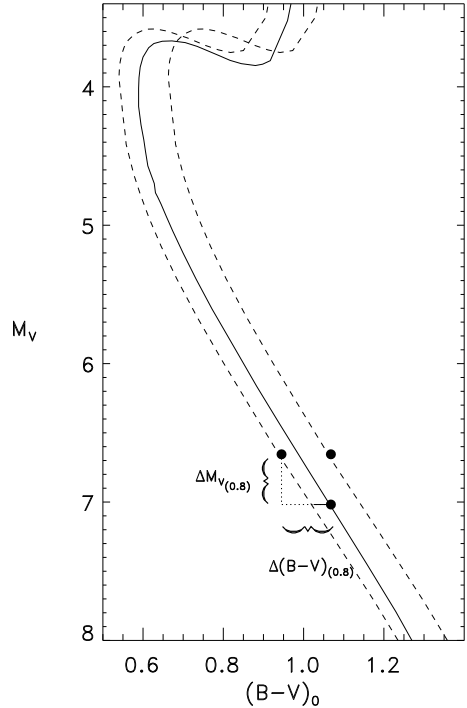

Fig. 7.— The relationship between the location of 0.8 $M_\odot$ ZAMS models for different metallicities and the color offset separating their parent isochrones. The 8 Gyr isochrones shown are [Fe/H] = 0.0 (dashed lines) and [Fe/H] = +0.15 (solid line). The filled circles represent the position of the 0.8 $M_\odot$ model on the unevolved main-sequence in each case. The quantities $\Delta M_{V(0.8)}$ and $\Delta(B-V)_{(0.8)}$ are the shifts in brightness and color between the 0.8 $M_\odot$ models. Only a fraction of $\Delta(B-V)_{(0.8)}$ (indicated by the filled-in portion of the dotted line) is required to register the main-sequence, turnoff and subgiant regions of the two isochrones. The redmost dashed line shows the significant misalignment resulting from a shift of the solar abundance isochrone by $\Delta(B-V)_{(0.8)}$ itself.

The $(B-V)_{(0.8)}$ vs. Z relation is insufficient by itself to determine the isochrone color shift with metallicity, since, for higher Z, the 0.8 $M_\odot$ ZAMS point becomes not only redder but also fainter. As a result, *the amount by which isochrones shift toward the red as the* *metallicity is increased is considerably less than the color difference between the corresponding 0.8 $M_\odot$ ZAMS points.* We illustrate this in Fig. 7 using two isochrones that differ only in metallicity (Z = 0.0169 vs. Z = 0.024). The offsets between the 0.8 $M_\odot$ ZAMS points (see Figs. 5 and 6 and Table 2) are marked by dotted lines, but the color shift needed to align the isochrones, $\delta(B-V)$, is only about one-third of $\Delta(B-V)_{(0.8)}$ as indicated by the filled-in portion of the line. As long as the slope of the lower main sequence does not change significantly, which seems to be the case for the metallicity and age range relevant here, we can derive values for $\delta(B-V)$ at fixed age as a function of Z. The resulting curve, determined using the values in Table 2 and a main sequence slope $\Delta(B-V)/\Delta M_V = 4.78$, is shown in Fig. 8.

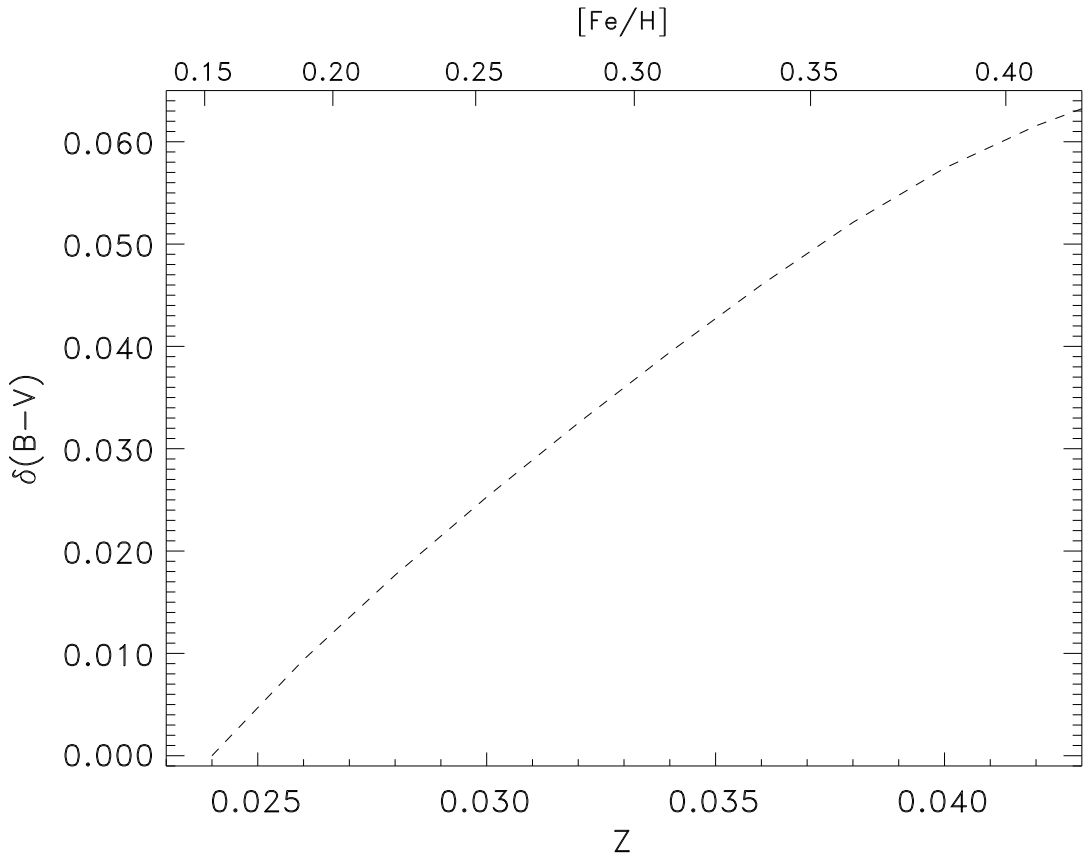

Fig. 8.— The quantity, $\delta(B-V)$, by which one can shift a Z = 0.024 isochrone to simulate progressively higher metallicities. It is measured in magnitudes and calculated from the behavior of 0.8 $M_\odot$ ZAMS models as explained in the text.

Finally, by subtracting the isochrone shifts ($\delta(B-V)$ from Fig. 8) from $\Delta(B-V)$ (0.25 and 0.275 for the KU92 and MJP94 data, respectively), we derive a pair of inverse relations between the cluster reddening and metallicity. These are indicated by the dashed lines in

Fig. 9. Each sequence begins with a point marking the limiting case where the measured $\Delta(B-V)$ is considered to be due entirely to reddening. The two filled circles represent the original spectroscopic determination of the cluster metallicity by Friel & Janes (1993) and the revised value quoted by Garnavich et al. (1994) based on a higher reddening value. However we have slightly updated these results for the present application. First, one of the nine stars turns out to be a nonmember (Liebert, Saffer & Green 1994) which causes the weighted mean for the remaining objects to rise to [Fe/H] = +0.23 (for E($B - V$) = 0.12). Second, we are concerned with the error of the mean rather than the scatter about the mean. This can be computed most simply as the standard deviation divided by the square-root of the number of measurements, which comes to 0.065 in this case. However, since Friel & Janes (1993) list the error for each object ($\Delta x_n$) we can more accurately compute

$$\frac{\sqrt{\sum (\Delta x_n)^2}}{N} = 0.085$$

We conservatively adopt an estimate for the error in the mean of ±0.09 and plot the spectroscopic results at E($B - V$) = 0.12, [Fe/H] = +0.23 and E($B - V$) = 0.19, [Fe/H] = +0.32 on Fig. 9. The intersection of this relation (and its error range) with our pair of relations (derived as explained above) encloses a roughly rectangular area in Fig. 9. For the KU92 data the results can be stated as E($B - V$) = 0.205±0.015 (i.e., $\delta(B-V)$ = 0.045), corresponding to [Fe/H] = +0.34 $\mp$ 0.07. If the slightly larger MJP94 color offset is used, the result would change to E($B - V$) = 0.225 ± 0.015 (i.e., $\delta(B-V)$ = 0.050) with [Fe/H] = +0.36 $\mp$ 0.07. Taking the systematic difference between the data sets as well as the uncertainty in the spectroscopy into account, we conclude from Fig. 9 that 0.19 < E($B - V$) < 0.24 and +0.44 > [Fe/H] > +0.27 for NGC 6791.

We can check our results for consistency by using E($B - V$) to compute the expected value of E($V - I$), subtracting it from the measured $\Delta(V - I)$ to find $\delta(V - I)$ and then comparing this with the expected behavior of $\delta(V - I)$ as a function of Z. The results of Dean, Warren & Cousins (1978) indicate that the color excess ratio for stars with colors near that of the cluster turnoff is E($V - I$)/E($B - V$) $\approx$ 1.3, so that for the KU92 data, E($V - I$) $\approx$ 0.27. Since $\Delta(V - I)_{KU92}$ = 0.315, we are left with $\delta(V - I)$ = 0.045 mag = $\delta(B-V)$. An analysis of the expected shift of an isochrone in ($V - I$) for small changes in metallicity (based on the behavior of 0.8 M$_\odot$ ZAMS models as was done for ($B-V$)) confirms that the two colors do, in fact, display nearly identical sensitivity to metallicity over the relevant color range.

If the metallicity of NGC 6791 is actually 0.2 dex higher in [Fe/H] than the isochrones which we fit, how will the derived age and distance be affected? Increasing metallicity will cause both the turnoff and the ZAHB to become fainter at a given age; the difference between the sensivities of the two will determine the functional dependence of $\Delta V_{TO}^{HB}$ on age. For the ages and metallicities of globular clusters, the turnoff is expected to fade by about 0.35 magnitudes per dex change in [Fe/H] (Buonanno, Corsi & Fusi Pecci 1989). The fading of the HB has been the subject of some controversy, but the slopes derived from various methods have ranged from 0.0 to 0.35, that is, $\Delta V_{TO}^{HB}$ at a given age slowly grows with increasing metallicity. In this case that means that the measured $\Delta V_{TO}^{HB}$ at [Fe/H] = +0.35 would correspond to an age younger (by perhaps 0.5 Gyr) than at +0.15. However, the metallicity range relevant here is considerably higher than that of globular clusters. Preliminary modelling indicates that the ZAHB does, in fact, fade more quickly as a function of Z at such high metallicities, but the turnoff may be expected to behave similarly. Consequently we believe that our age result can best be stated as $10 \pm 0.5$ Gyr.

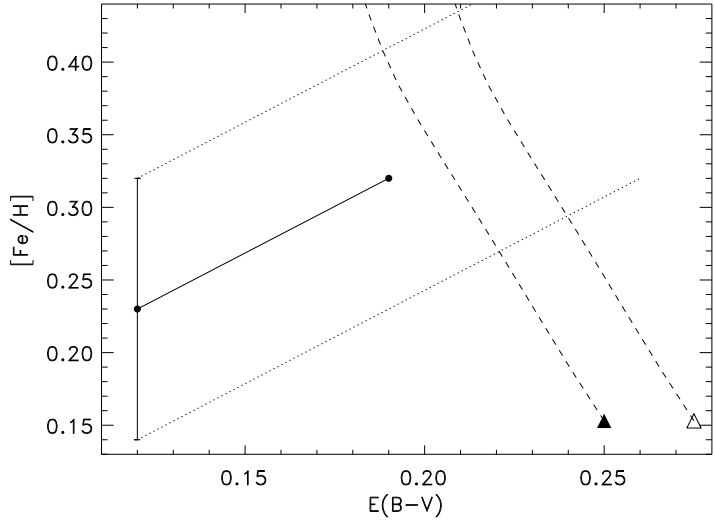

Fig. 9.— Range of consistent results for the cluster metallicity and foreground reddening based both on spectroscopy and isochrone fitting. The filled circles represent the weighted mean metallicity for eight NGC 6791 giants whose spectra were analysed by Friel & Janes (1993) at two different assumed reddening values. Error bars indicate the computed error in the mean; these are extended to maintain the error range as the derived mean metallicity *increases* with reddening. The dashed lines show the result of fitting our 10 Gyr isochrone for [Fe/H] = +0.15 to the cluster CMDs, assuming that the total color offset needed to fit each results from a small metallicity deficit in addition to the actual foreground reddening. In this case the derived cluster metallicity *decreases* with increasing reddening. The filled triangle marks the measured $\Delta(B-V)$ for the KU92 data, the open triangle MJP94. The intersection of these relations defines the locus of consistent metallicity and reddening results.

The derived apparent distance modulus will simply increase in lockstep with the HB as it fades with increasing metallicity, an effect which our preliminary analysis indicates may be as much as 0.18 magnitudes as Z changes from 0.024 to 0.040. Pending more detailed analysis, we estimate that $13.49 \lesssim (m - M)_V \lesssim 13.70$.

### 3.3. Clump Star Mass

In TDB93, we used solar metallicity isochrones plus evolutionary tracks for the subsequent He-burning phases to determine the mass of the clump giant stars in M67. We found evidence for a substantial amount of mass loss having taken place, such that the RGB stars (with mass $\approx 1.27 M_\odot$) become clump stars with $M \lesssim 0.7 M_\odot$. The excellent agreement between the NGC 6791 photometry and the theoretical isochrones and ZAHB shown in Figs. 1 and 2 suggests that the clump stars in this cluster have nearly the same mass as those in M67. The implied amount of mass lost, however, is not as great in this case since the NGC 6791 giants have somewhat lower masses than M67 giants, ($\approx 1.10 M_\odot$) primarily because of their greater age.

It is interesting to compare the predictions of the Reimers mass-loss formula (Reimers 1975) with these results. The simplest way to make this comparison is in terms of the parameter, $\eta$, needed to fit the observations via the Reimers formula. Renzini (1981) has shown that the morphology of globular cluster horizontal-branches constrains $\eta = 0.4 \pm 0.2$. On the other hand, Fig. 4 of TDB93 implies that for M67, $\eta \approx 2.0$. Applying the same analysis to the present data yields the result that the mass of the NGC 6791 clump stars are consistent with the Reimers formula if $\eta \approx 1.0$. This finding runs counter to the hypothesis of TDB93 that, above some threshold, mass-loss increases with metallicity, since NGC 6791 is approximately twice as metal-rich as M67. There is an apparent age dependence in the sense that $\eta$ is greatest in the youngest object considered here, M67, and is lowest in the globular clusters. This would be most sensibly recast as an dependence on RGB mass. However there is no theoretical justification for such an effect, as discussed by TDB93, and it is, in fact, contrary to the sense of the Reimers law and any of the several parametrized mass-loss relations which have been advanced. We continue to speculate, then, that mass-loss on the first-ascent RGB acts to reduce the size of the stellar envelope *to* rather than *by* a constant amount of mass.

We may once again ask how an increase in [Fe/H] over the +0.15 of the isochrones, as suggested in the previous section, affects our conclusions regarding mass loss. First, the theoretical ZAHB will edge closer to the RGB if Z increases with all else remaining constant. We have computed that this amounts to a temperature difference of only about 20

K, however, and may be safely neglected. In any case it goes in the direction of decreasing the derived clump star masses. Second, at higher metallicity the RGB stars will be higher in mass at a given age (by about 0.07 M$_\odot$ for $Z = 0.024$ vs. $Z = 0.040$). But the Reimers formula applied to RGB tracks implies slightly greater mass loss for the higher metallicities (as the stellar radii are larger), which cancels the effect. Thus, there should be little or no impact on our mass-loss results from a small metallicity increase.

## 4. Comparison with Previous Studies

NGC 6791 has been the object of quite a number of studies over the past thirty years, yet there has been little agreement on the basic cluster parameters. Estimates of the age have ranged over a factor of two, from 6 Gyr (Anthony-Twarog & Twarog 1985) to 12.5 Gyr (Janes 1988, Kałużny 1990). Our result of $10 \pm 0.5$ Gyr (based on fitting both the main-sequence turnoff region *and* the He-burning clump stars) falls within those boundaries and confirms the identification of NGC 6791 as one of the oldest known open clusters (Phelps, Janes & Montgomery 1994). This result is also consistent with the age estimate of the Galactic disk derived from the luminosity of the faintest white dwarfs and the sum of their cooling times and prior evolutionary times by Wood (1992).

Determinations of the cluster reddening have tended to fall into two groups (as noted by Liebert, Saffer, & Green 1994): either low values near E$(B-V) = 0.10$ (eg, Harris & Canterna 1981, Janes 1984, MJP94) or else high values around 0.20 (eg, Kinman 1965, Anthony-Twarog & Twarog 1985, KU92). Our range of results, $0.19 <$ E$(B-V) < 0.24$, is consistent with the latter group and argues strongly against values as low as those proposed by the former.

## 5. Summary

We have computed new isochrones for ages 8, 10 and 12 Gyr at a metallicity of [Fe/H] $\approx$ +0.15 (i.e., Z = 0.024). In addition we have generated a model zero-age horizontal branch for stellar masses between 0.55 and $1.3 M_\odot$ using identical physics, boundary conditions, etc. These have been compared to two independent sets of photometric data for stars in the central field of the rich open cluster NGC 6791. The comparisons of theory to observation yield the following results:

(1) Our 10 Gyr isochrone provides an excellent fit to the Kałużny & Udalski (1992) CCD photometry in the $M_V$ vs. $(B-V)$ and $M_V$ vs. $(V-I)$ planes if color shifts of 0.25 and 0.315 magnitudes, respectively, are applied. The data from Montgomery, Janes & Phelps (1994) are systematically a bit redder; fitting these requires shifts of 0.275 and 0.37 mag. The apparent distance modulus to NGC 6791 is $13.49 < (m-M)_V < 13.70$, depending on the cluster metallicity and the uncertainty in HB brightness as Z increases. The age determination is reinforced by the excellent fit simultaneously to the red clump stars and main-sequence turnoff; we conclude that NGC 6791 has an age of $10 \pm 0.5$ Gyr.

(2) The size of the color offsets $\Delta(B-V)$ and $\Delta(V-I)$ when combined with the spectroscopic results of Friel & Janes (1993) imply that the foreground reddening to NGC 6791 lies in the range $0.19 < E(B-V) < 0.24$ with a coupled metallicity range of $+0.44 >$ [Fe/H] $> +0.27$.

(3) The zero-age horizontal branch models suggest that the clump stars in NGC 6791 have masses of $\lesssim 0.7 M_\odot$. A similar result was found by Tripicco, Dorman & Bell (1993) for the M67 clump stars, whose red giant predecessors are nearly $0.2 M_\odot$ heavier. This, combined with masses derived for globular cluster horizontal branch stars, suggests the presence of a mechanism which reduces stellar envelopes *to* a given size rather than *by* the amount predicted by any of the commonly used mass-loss parameterization schemes.

This research was supported by NSF grants AST-9122361 and AST-9314931, and by NASA Long Term Space Astrophysics Research Program grant NAGW-2596. Computing support was provided to B.H. by the National Radio Astronomy Observatory.

---